\documentclass[conference]{IEEEtran}
\IEEEoverridecommandlockouts

\usepackage{cite}
\usepackage{amsmath,amssymb,amsfonts}
\usepackage{algorithmic}
\usepackage{graphicx}
\usepackage{textcomp}
\usepackage{xcolor}
\usepackage{etoolbox}
\def\BibTeX{{\rm B\kern-.05em{\sc i\kern-.025em b}\kern-.08em T\kern-.1667em\lower.7ex\hbox{E}\kern-.125emX}}

\graphicspath{ {./fig/} }
\usepackage{siunitx}
\usepackage{orcidlink}
\usepackage{enumitem}
\usepackage{pgfplots}
\pgfplotsset{compat=1.18}
\usepackage{multirow}
\usetikzlibrary{positioning, arrows.meta}
\usetikzlibrary{calc}
\usepackage[caption=false, font=footnotesize]{subfig}
\DeclareRobustCommand{\tikzmarker}[1]{%
	\:%
	\tikz\draw[black,mark=#1,mark options={scale=1,fill=black}] plot coordinates {(0,0)};
}
\usepackage[final]{changes}
\usepackage{pbalance}

\definecolor{red191000}{RGB}{255,0,0}				
\definecolor{goldenrod1911910}{RGB}{191,191,0}		
\definecolor{gray127127127}{RGB}{127,127,127}		

\definecolor{darkviolet1910191}{RGB}{191,0,191}		
\definecolor{green01270}{RGB}{0,127,0}				
\definecolor{teal027127127}{RGB}{27,127,127}		

\definecolor{darkturquoise0191191}{RGB}{0,191,191}	
\definecolor{black000000}{RGB}{0,0,0}				

\definecolor{blue00191}{RGB}{0,0,227}				
\definecolor{orange2551270}{RGB}{255,127,0}			

\definecolor{lightred255153153}{RGB}{204,102,102}		
\definecolor{lightgoldenrod255255153}{RGB}{204,204,102}	
\definecolor{lightblue153204255}{RGB}{153,153,153}		

\definecolor{darklavender191102191}{RGB}{191,102,191}	
\definecolor{darkgreen102204102}{RGB}{102,204,102}		
\definecolor{darkpeach204153102}{RGB}{204,153,102}		

\definecolor{darkgray176}{RGB}{176,176,176}
\definecolor{lightgray204}{RGB}{204,204,204}

\definecolor{darkblue102153204}{RGB}{102,153,204}           
\definecolor{darkpeach204153102}{RGB}{204,153,102}          

\newcommand{\treeCase}[1]{
	\begin{tikzpicture}
		\begin{semilogxaxis}[
			width=1\columnwidth,
			height=6.5cm,
			legend cell align={left},
			legend style={
				fill opacity=0.8,
				draw opacity=1,
				text opacity=1,
				at={(0.02,0.98)},
				anchor=north west,
				draw=lightgray204
			},
			log basis x={10},
			tick align=outside,
			tick pos=left,
			x grid style={darkgray176},
			xlabel={Number of keys in tree},
			xmajorgrids,
			xmin=0.446683592150963, xmax=22387211.3856834,
			xmode=log,
			minor x tick num=8,
			major tick length=4pt,
			minor tick length=2pt,
			xtick style={color=black},
			minor tick style={color=black},
			y grid style={darkgray176},
			scaled y ticks=false,
			yticklabel style={/pgf/number format/fixed},
			ylabel={Time in \si{\micro\second}},
			ylabel style={yshift=-0.5ex},
			ymajorgrids,
			ymin=-540.95677499953, ymax=15000,
			ytick style={color=black},
			ytick={0, 2500, 5000, 7500, 10000, 12500},
			clip=false,
			]
			#1
		\end{semilogxaxis}
	\end{tikzpicture}
}

\newcommand{\batchCase}[1]{
	\begin{tikzpicture}
		\begin{axis}[
			width=1\columnwidth,
			height=6cm,
			legend cell align={left},
			legend style={
				fill opacity=0.8,
				draw opacity=1,
				text opacity=1,
				at={(0.02,0.98)},
				anchor=north west,
				draw=lightgray204
			},
			tick align=outside,
			tick pos=left,
			x grid style={darkgray176},
			xlabel={Batch Size},
			xmajorgrids,
			xmin=-45.8, xmax=1049.8,
			xtick style={color=black},
			y grid style={darkgray176},
			scaled y ticks=false,
			yticklabel style={/pgf/number format/fixed},
			ylabel={Time in \si{\micro\second}},
			ylabel style={yshift=-0.5ex},
			ymajorgrids,
			ymin=-125, ymax=4000,
			ytick style={color=black},
			ytick={0, 1000, 2000, 3000},
			clip=false,
			]
			#1
		\end{axis}
	\end{tikzpicture}
}

\newcommand{\batchCaseSmallCU}[1]{
	\begin{tikzpicture}
		\begin{axis}[
			width=1\columnwidth,
			height=6.5cm,
			legend cell align={left},
			legend style={
				fill opacity=0.8,
				draw opacity=1,
				text opacity=1,
				at={(0.02,0.98)},
				anchor=north west,
				draw=lightgray204
			},
			tick align=outside,
			tick pos=left,
			x grid style={darkgray176},
			xlabel={Batch Size},
			xmajorgrids,
			xmin=-45.8, xmax=1049.8,
			xtick style={color=black},
			y grid style={darkgray176},
			scaled y ticks=false,
			yticklabel style={/pgf/number format/fixed},
			ylabel={Time in \si{\micro\second}},
			ylabel style={yshift=-0.5ex},
			ymajorgrids,
			ymin=-75, ymax=2400,
			ytick style={color=black},
			ytick={0, 500, 1000, 1500, 2000},
			clip=false,
			]
			#1
		\end{axis}
	\end{tikzpicture}
}

\newcommand{\batchCaseCPU}[1]{
	\begin{tikzpicture}
		\begin{axis}[
			width=1\columnwidth,
			height=5cm,
			legend cell align={left},
			legend style={
				fill opacity=0.8,
				draw opacity=1,
				text opacity=1,
				at={(0.02,0.98)},
				anchor=north west,
				draw=lightgray204
			},
			tick align=outside,
			tick pos=left,
			x grid style={darkgray176},
			xlabel={Batch Size},
			xmajorgrids,
			xmin=-45.8, xmax=1049.8,
			xtick style={color=black},
			y grid style={darkgray176},
			scaled y ticks=false,
			yticklabel style={/pgf/number format/fixed},
			ylabel={Time in \si{\micro\second}},
			ylabel style={yshift=-0.5ex},
			ymajorgrids,
			ymin=-639.8125, ymax=17214.5625,
			ytick style={color=black},
			clip=false,
			]
			#1
		\end{axis}
	\end{tikzpicture}
}

\newcommand{\batchCaseCPUThread}[1]{
	\begin{tikzpicture}
		\begin{axis}[
			width=1\columnwidth,
			height=5.5cm,
			legend cell align={left},
			legend style={
				fill opacity=0.8,
				draw opacity=1,
				text opacity=1,
				at={(0.02,0.98)},
				anchor=north west,
				draw=lightgray204
			},
			tick align=outside,
			tick pos=left,
			x grid style={darkgray176},
			xlabel={Batch Size},
			xmajorgrids,
			xmin=-45.8, xmax=1049.8,
			xtick style={color=black},
			y grid style={darkgray176},
			scaled y ticks=false,
			yticklabel style={/pgf/number format/fixed},
			ylabel={Time in \si{\micro\second}},
			ylabel style={yshift=-0.5ex},
			ymajorgrids,
			ymin=-350, ymax=10000,
			ytick style={color=black},
			clip=false,
			]
			#1
		\end{axis}
	\end{tikzpicture}
}

\begin{document}

\title{
	Efficient Batch Search Algorithm for B+ Tree Index Structures with Level-Wise Traversal on FPGAs\\
}

\author{
	\ifbool{true}{
		\IEEEauthorblockN{
			Max Tzschoppe\,\orcidlink{0009-0002-9761-5994}\,\IEEEauthorrefmark{1},
			Martin Wilhelm\,\orcidlink{0000-0002-2713-2842}\,\IEEEauthorrefmark{1},
			Sven Groppe\,\orcidlink{0000-0001-5196-1117}\,\IEEEauthorrefmark{2},
			Thilo Pionteck\,\orcidlink{0000-0001-6518-1226}\,\IEEEauthorrefmark{1}
		}
		\IEEEauthorblockA{\IEEEauthorrefmark{1} \textit{Institute for Information Technology and Communications, Otto-von-Guericke University Magdeburg, Germany}}
		\IEEEauthorblockA{\IEEEauthorrefmark{2} \textit{Institute of Computer Science, TU Bergakademie Freiberg, Germany}}
	}{
		\IEEEauthorblockN{
			double blinded
		}
		\IEEEauthorblockA{\IEEEauthorrefmark{0} \textit{}}
		\IEEEauthorblockA{\IEEEauthorrefmark{0} \textit{}}
	}
}

\maketitle


\begin{abstract}

	This paper introduces a search algorithm for index structures based on a B+ tree, specifically optimized for execution on a field-programmable gate array (FPGA). Our implementation efficiently traverses and reuses tree nodes by processing a batch of search keys level by level. This approach reduces costly global memory accesses, improves reuse of loaded B+ tree nodes, and enables parallel search key comparisons directly on the FPGA. Using a high-level synthesis (HLS) approach, we developed a highly flexible and configurable search kernel design supporting variable batch sizes, customizable node sizes, and arbitrary tree depths. The final design was implemented on an AMD Alveo U250 Data Center Accelerator Card, and was evaluated against the B+ tree search algorithm from the TLX library running on an AMD EPYC 7542 processor (2.9$\,$GHz). With a batch size of 1000 search keys, a B+ tree containing one million entries, and a tree order of 16, we measured a 4.9$\times$ speedup for the single-kernel FPGA design compared to a single-threaded CPU implementation. Running four kernel instances in parallel on the FPGA resulted in a 2.1$\times$ performance improvement over a CPU implementation using 16 threads.
\end{abstract}


\section{Introduction} \label{sec:introduction}

\added{The growing demand for high-throughput analysis of large-scale datasets poses significant challenges to core components of modern database systems. One essential aspect is main memory capacity, which has grown over the last decade to several terabytes in today's servers. As a result, many database systems now operate entirely in-memory~\cite{Kabakus2017}, giving rise to specialized in-memory database (IMDB) systems that drastically reduce access latencies.} Index structures are a fundamental component of these modern IMDB systems. They store search keys together with pointers to the corresponding data records, enabling queries to be executed efficiently without scanning entire tables. Consequently, the performance of index structures is critical for overall database performance and has increasingly become a primary system bottleneck~\cite{Kocberber2013}.

Index structures support not only search operations but also insertions and deletions. Nevertheless, search performance is the dominant factor for query efficiency, particularly in the context of data warehouses, which form the backbone for managing diverse and interconnected data sources. This is especially evident in online analytical processing (OLAP) workloads, which consist mainly of read-intensive SQL queries~\cite{Jarke2003}. OLAP workloads are central to business intelligence applications, where insights are derived from large volumes of historical data. In these scenarios, queries often focus on a relatively small hot subset of the dataset. Such subsets are frequently cached in order to reduce query latency and system load~\cite{Nicholson2023}. Since these subsets typically represent historical data that changes infrequently, the cached structures remain mostly static, require only occasional updates, and are commonly implemented using index structures.

Another important scenario involving large volumes of predominantly static data is found in machine learning (ML) and large language model training pipelines. These pipelines often rely on aggregated data originating from data warehouses~\cite{Bodner2025}, which necessitates efficient access to vast amounts of structured and semi-structured data. As the scale of such workloads continues to grow, further improvements in query latency and throughput become increasingly important.

Tree-based index structures, such as B-trees and B+ trees, are widely used in database systems due to their efficient search performance, which is characterized by logarithmic time complexity~\cite{Schneider2004}. In particular, the B+ tree benefits from variable node sizes, which reduce tree depth and improve lookup efficiency. Its balanced structure and the organization of data exclusively at the leaf level also enable options for hardware acceleration.

Optimized index structures based on B+ trees have been developed for CPU execution~\cite{Liu2024} as well as for offloading tree operations to hardware accelerators such as FPGAs~\cite{Melikoglu2019, Heinrich2015, Fang2019} and GPUs~\cite{Shahvarani2016, Fang2019}. On CPUs, query processing typically involves individual lookups per request combined with a linear or binary search within each tree node, which limits throughput and parallelism. In contrast, hardware accelerators lack the high clock frequencies and advanced caching hierarchies of CPUs but they offer the advantage of massive parallelism and customizable data paths. GPUs and \mbox{FPGAs} therefore offer significant potential for accelerating such workloads by leveraging their inherent parallelism and enabling efficient memory access strategies. Accordingly, this work focuses on the design and implementation of a novel FPGA-optimized search algorithm for B+ trees.
%
%
Our design loads individual nodes from the DDR memory of the FPGA card and processes all corresponding search queries simultaneously to minimize costly memory accesses. Comparisons are performed in parallel across all key entries within a node, utilizing a custom comparison logic based on a Cascading Bitwise Priority Comparison (CBPC) combined with a priority encoder. For evaluation purposes, we developed a fully automated testing framework focused on comprehensive and reliable assessments. To ensure an unbiased workload, both the search keys and the tree entries are randomly generated. We\added{ evaluated the design on the AMD Alveo U250 Data Center Accelerator Card and} compared the results against single- and multi-threaded CPU execution to ensure a realistic test scenario.

To the best of our knowledge, no prior work has implemented or analyzed a complete B+ tree search algorithm on an FPGA. The proposed design supports fully flexible tree sizes, constrained only by the available DDR memory, and allows an arbitrary batch size (the number of search keys) up to a predefined maximum. A feature of this approach is that both parameters, tree size and batch size, can be adjusted at runtime. Since the comparison and buffering logic depend on the tree order (the maximum number of key entries per node), we evaluate multiple design variants with fixed tree orders to enable a comprehensive design space exploration.

The remainder of this paper is structured as follows. Sec.~\ref{sec:related_work} reviews related work on tree acceleration using FPGAs and GPUs. Sec.~\ref{sec:background} provides essential background on the B+ tree structure. Sec.~\ref{sec:design} introduces our FPGA-optimized B+ tree search algorithm and describes the proposed design in detail. Sec.~\ref{sec:evaluation} presents a comprehensive evaluation of our approach, including various measurement results. Finally, Sec.~\ref{sec:conclusion} concludes the paper with a summary of the key findings\deleted{ and outlines directions for future work}.

\section{Related Work} \label{sec:related_work}

Trees are indispensable for implementing efficient index structures. The following related work section focuses on how tree operations can be accelerated using GPUs or FPGAs.

\textbf{Tree Operation Acceleration on FPGAs.}
Several research groups have investigated the use of FPGAs to accelerate B-tree operations. Liu et al.~\cite{Liu2024} present the Honeycomb system, which accelerates get and scan operations on an FPGA. They report a throughput improvement of 2$\times$ for scan-heavy workloads compared to a multi-threaded CPU implementation. This particular workload is well-suited for FPGA execution. This means that the observed performance gains may not be applicable to more realistic workloads.
Melikoglu et al.~\cite{Melikoglu2019} implemented an optimized B-tree search algorithm on an FPGA and compared it with a non-optimized baseline version, achieving a throughput gain of 8$\times$. The study lacks a comparison against CPU-based search performance and is primarily relevant in the context of hardware architecture research. Heinrich et al.~\cite{Heinrich2015} proposed a hybrid approach in which a B+ tree is distributed between the CPU and FPGA. Their design accelerates the upper levels of the tree on the FPGA while completing the search on the CPU. They report a maximum speedup of 2.3$\times$ compared to pure CPU execution. However, accelerating only the upper tree levels can become a bottleneck for very large trees, where the number of nodes per level increases significantly in the lower levels.

\textbf{Tree operation acceleration on GPUs.}
In addition to FPGA-based approaches, several works have investigated GPU acceleration for tree operations. Shahvarani et al.~\cite{Shahvarani2016} proposed a hybrid design in which a B+ tree is partitioned between the CPU and GPU, with the lower tree levels processed on the GPU. The authors report a speedup of 2.4$\times$ over a CPU-only execution. Their comparison is made against a single-threaded CPU implementation, whereas modern multi-threaded CPU-based B+ tree designs would likely yield significantly better performance.
Another notable example of GPU-based acceleration is presented by Koppehel et al.~\cite{Koppehel2021}, who introduce CuART, a scalable and parallel engine for lookup and update operations on adaptive radix trees, implemented in CUDA. The design leverages warp-level parallelism and\deleted{ optimized} memory access strategies to maximize throughput. CuART achieves up to a 5$\times$ speedup for mixed workloads compared to a single-threaded CPU baseline. However, similar to previous works, it lacks a direct comparison against optimized multi-threaded CPU tree implementations.

Overall, the presented related works on accelerating tree operations using FPGAs and GPUs demonstrates promising speedups but often relies on limited or unbalanced comparisons. Many studies evaluate performance against single-threaded CPU baselines rather than modern multi-threaded implementations, reducing the validity of claimed gains. Additionally, several FPGA-based approaches only accelerate upper tree levels or rely on host memory, introducing bottlenecks that limit scalability for large tree structures.

With our proposed FPGA design, we address these limitations by implementing a complete B+ tree search algorithm over all tree levels that supports fully flexible tree sizes. Unlike prior work, we perform a fair and comprehensive comparison against multi-threaded CPU implementations to assess realistic performance gains.

\section{B+ Tree Background} \label{sec:background}

\begin{figure}[h!]
	\centering
	\includegraphics[width=0.65\columnwidth]{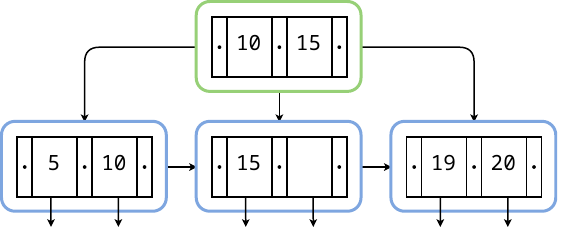}
	\caption{Simple and small B+ tree structure with order $m=3$.}
	\label{fig:btree_example}
	\vspace{-5pt}
\end{figure}

The B+ tree is a widely adopted index structure used in both open-source~\cite{MySQL2025, PostgreSQL2025} and commercial~\cite{IBM2025, Oracle2025} databases and file systems, supporting efficient query operations on large, sorted datasets~\cite{Schneider2004}. The concept of B-trees was originally developed by Bayer et al.~\cite{Bayer1972}, with the B+ tree being an extension that organizes all data entries at the leaf level, while internal nodes contain only keys. Two key characteristics define the B+ tree:
\begin{enumerate}
	\item Its balanced tree structure ensures consistent key lookup times with logarithmic time complexity.
	\item Its adjustable node size allows for a high fan-out, which minimizes the height of the tree.
\end{enumerate}
\noindent
In addition for file-intensive operations, the B+ tree’s structure, where all leaf nodes are linked in a linear sequence, enables efficient sequential access from the leftmost to the rightmost leaf node. A small example of a B+ tree is shown in Fig.~\ref{fig:btree_example}, where the green node represents an internal node and the blue nodes correspond to leaf nodes.
%

\textbf{B+ Tree Metrics.} The order of a B+ tree, denoted by $m$, defines the maximum number of child pointers per inner node. Consequently, the maximum number of keys $k_{\text{max}}$ in an inner node is $k_{\text{max}} = m - 1$. To maintain the balanced structure of the tree, each internal node must have at least $\frac{m}{2}$ child pointers, meaning it must be at least half full. To estimate the size of a full B+ tree, the maximum number of nodes $N_{\text{max}}$ can be calculated as: $N_{\text{max}} = \sum_{i=0}^{h-1} m^i$ where $h$ denotes the height (i.e., the number of levels) of the tree. Another important metric is the maximum number of keys per tree level, given by $L_{\text{max}} = m^h \cdot k_{\text{max}} = m^h \cdot (m-1)$. This value indicates how many keys can be stored at a given level before the tree must grow in height by adding a new level.


\section{FPGA-Optimized B+ Tree Search Algorithm} \label{sec:design}

The following section presents our FPGA-optimized B+ tree search algorithm, outlining the core concept, the memory organization, and the architecture of the search procedure.

\subsection{Idea}

Conventionally, multiple search queries are processed sequentially, with each query starting after the previous one has completed. In contrast, our approach reinvents this concept\replaced{ by collecting queries and processing them in large batches}{ and implements it using a high-level synthesis (HLS) workflow}.

The HLS design was kept as simple and modular as possible. The development of an efficient implementation is guided by two key principles. First, the design should process as many steps of the algorithm in parallel as possible to take advantage of the hardware design. Second, memory accesses should be minimized, as they are costly in terms of latency.

\begin{figure}[]
	\centering
	\includegraphics[width=0.45\textwidth]{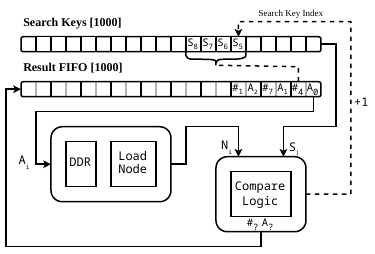}
	\caption{Search Algorithm Overview.}
	\label{fig:search_algorithm_overview}
	\vspace{-5pt}
\end{figure}

Fig.~\ref{fig:search_algorithm_overview} provides an overview of the proposed implementation strategy. The core idea is to process queries in batches of search keys \texttt{S} by traversing the B+ tree level by level, while storing intermediate results in a FIFO buffer. Each entry in the result FIFO contains two fields: the address \texttt{A} of a node and the number \texttt{\#} of search keys that need to be compared with that node. The general flow of our search algorithm proceeds as follows: a node \texttt{N} is loaded from the address specified in the first FIFO entry. Once the node is available, it is forwarded to the comparison logic, where the corresponding number of search keys \texttt{\#} are sequentially compared against the search keys \texttt{S} defined by them. Each comparison is done in parallel and produces a new node address and a number of associated search keys.
%

Fig.~\ref{fig:search_algorithm_overview} illustrates how node \texttt{N$_\text{0}$} is loaded from address \texttt{A$_\text{0}$}. The label \texttt{\#$_\text{4}$} indicates that four search keys need to be processed for this node. As a result, node \texttt{N$_\text{0}$} is compared with search keys \texttt{S$_\text{5}$} through \texttt{S$_\text{8}$}. During each iteration, the \texttt{SearchKeyIndex} is incremented, and the comparison results are written to the result FIFO. This approach preserves the processing order and simplifies access to the search keys.

Since the batch of queries is sorted, progressively smaller sub-batches are passed to each subsequent level during successive memory access. This approach minimizes overhead by ensuring that all queries targeting the same node are processed in a single memory access. In summary, the entire query batch is processed level by level, loading only the nodes required for the current set, and this enables an efficient level-wise traversal strategy. However, the benefit diminishes for higher tree levels due to the fewer queries per node if the queries are uniformly distributed.

\subsection{B+ Tree Memory Organization}

On the CPU, the B+ tree resides in main memory, and its nodes are connected via pointers. To transfer this structure to an FPGA, a mapper is required to transform the hierarchical tree into a flat array representation. This transformation is performed using a breadth-first search (BFS) algorithm. The resulting flat tree array includes both internal and leaf nodes. To ensure a simplified and uniform memory layout, all nodes are padded to have the same size. We chose to embed the child addresses directly within each node to avoid additional address computation on the FPGA. 
\begin{figure}[h!]
	\centering
	\includegraphics[width=1\columnwidth]{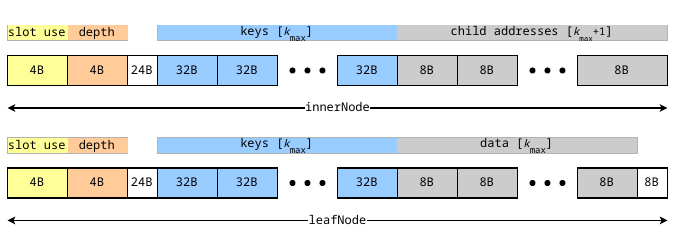}
	\caption{Node Layout}
	\label{fig:node_layout}
	\vspace{-5pt}
\end{figure}

The node layout for inner and leaf nodes are illustrated in Fig.~\ref{fig:node_layout}. Each node contains the number of active entries (\texttt{slotUse}), the level at which the node is located (\texttt{depth}) and node keys (\texttt{key}). The inner node ends up with the addresses of the child nodes (\texttt{childAddress}). The leaf node contains the actual data (\texttt{data}), with the last 8-bytes are left unused to maintain the same node size as an inner node. The sizes of the \texttt{key}, \texttt{childAddress} and \texttt{data} sections depend on the maximum number of key entries per node, denoted as $k_{\text{max}}$. The node size $N_{\text{size}}$ depends on the tree order $m$ as described in Sec.~\ref{sec:background} and can be calculated as:
\begin{align}
	N_{\text{size}} &= \SI{32}{\byte} + \SI{32}{\byte} \cdot k_{\text{max}} + \SI{32}{\byte} \cdot \frac{k_{\text{max}} + 1}{4} \label{eq:node_size}\\
	&= \SI{40}{\byte} \cdot (k_{\text{max}} + 1) = \SI{40}{\byte} \cdot m \nonumber
\end{align}

To maximize the data throughput between host and FPGA, the array of nodes is partitioned into 32-byte chunks. This uniform sizing enables a simplified memory access module and loading nodes from memory becomes more efficient and implementation-friendly.

\subsection{Search Key Memory Organization}

Search queries are collected by the host and transferred in batches to the memory of the FPGA board. Each batch consists of a 32-byte search key. Since the search keys are required during every iteration of the search process, a buffer with a fixed maximum length is implemented in the design. Before the search process begins, the entire batch is preloaded from memory into the buffer implemented using BRAM. This significantly reduces the access latency for search keys during execution. To avoid array partitioning and to stay within the available BRAM resources of the FPGA, the buffer size is set to accommodate 1000 entries. 


\subsection{Result Memory Organization}

To enable batch processing of search queries, the B+ tree is traversed level by level. This approach requires that intermediate results from each level are stored in a result FIFO. This FIFO serves a dual purpose. First, it temporarily holds intermediate results, which include the child address for the next node and the number of search keys assigned to that specific child node. Second it stores the final search results.

Thanks to the uniform structure of inner and leaf nodes, managing this process is straightforward. While processing internal nodes, 8-byte child address entries and an additional 4 bytes for the number of associated search keys are stored. When the traversal reaches the leaf level, the result format switches: each entry is either a -1 (indicating the search key was not found) or a 8-byte data value if a match is found—both of which are written to the same FIFO. The size of the FIFO is determined by the batch size.

\subsection{Parallel Key Comparison} \label{subsec:comparison}

The design's main component is the comparison module, which is responsible for comparing search queries with the currently loaded node. When a node is transferred from the memory to the FPGA, all search keys associated with that node are processed together. This coordination is managed by the result FIFO, which provides the number of keys to be compared with the node. This is feasible because the search keys are sorted and stored in BRAM, allowing for a simple step-by-step traversal of all search terms at each level.
\begin{figure}[h!]
	\centering
	\includegraphics[width=1\columnwidth]{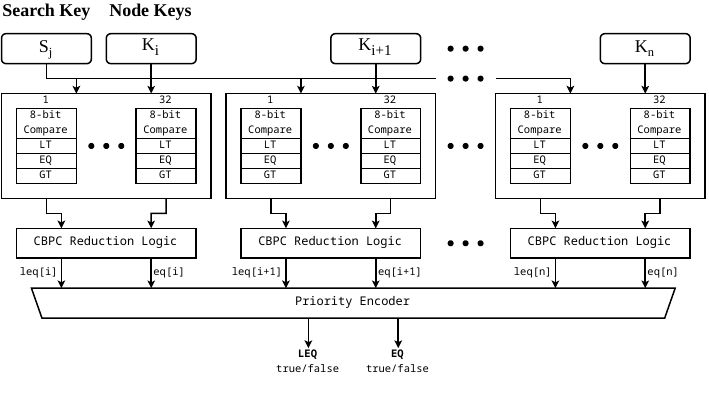}
	\caption{Key Comparison Procedure.}
	\label{fig:comparison}
	\vspace{-5pt}
\end{figure}

The flow of the key comparisons is depicted in Fig.~\ref{fig:comparison}. The number of key comparisons per node is determined by $k_{\text{max}}$, which must be specified prior to synthesis. This implies that each search key must be compared against all $k_{\text{max}}$ entries in the node to determine the correct child node address. Each individual comparison involves comparing a 32-byte search key with a 32-byte node key. This is implemented using 32 parallel 8-bit comparators, each producing a 3-bit result indicating whether the corresponding byte is less than, equal to, or greater than its counterpart. For each comparison, only one of these three bits can be active.

To determine which child node a search key should be routed to, it is sufficient to evaluate whether the key is less than or equal to the node entry, with equality also being used for identifying exact matches in the leaf nodes. Therefore, the 32 per-byte comparison results are reduced to a single comparison outcome using a Cascading Bitwise Priority Comparison (CBPC) unit. The CBPC is composed entirely of combinatorial logic and resolves the 32 intermediate results in a single step into one final comparison result.

This 32-byte comparison is performed in parallel for each of the $k_{\text{max}}$ slots in the node. Consequently, $k_{\text{max}}$ comparison outcomes (each indicating less/equal or equal) are generated. These are further reduced to a single final decision using a priority encoder. The combination of parallel 8-bit comparators, the CBPC, and the priority encoder encourages the HLS tool to synthesize highly parallel and efficient comparison logic.

\subsection{Search Algorithm}

The execution of the entire search algorithm can be described as follows: Before the actual search begins, all search keys are transferred from\added{ global} memory into the FPGA’s BRAM. Subsequently, a node—initially the root node—is loaded from memory. These load operations utilize burst reads to fully exploit the available memory bandwidth.

Each search key is then compared in parallel with the keys of the root node, as described in Sec.~\ref{subsec:comparison}. Every comparison yields the child address of the node to which the search key must be passed next. If the same child address corresponds to multiple search keys, a counter is incremented accordingly. This counter ultimately reflects the number of search keys associated with that specific child node. Both the child address and the corresponding count are stored in the result FIFO.

After processing the root node, the next iteration begins with the second level of the tree. Based on the first child address stored in the result FIFO, the corresponding next node is loaded, and the procedure is repeated until the leaf level is reached.

At the leaf level, the result FIFO no longer contains child addresses but instead holds the actual results or data entries corresponding to the search queries. In the final step, these results are collected and written back to the\added{ global} memory using burst writes so that they can be further\added{ accessed} and processed by the host.

%

\subsection{Kernel Parallelism}

\begin{figure}[h!]
	\vspace{-5pt}
	\centering
	\subfloat[Single Instance\label{subfig:single}]{\raisebox{0.7\height}{%
		\includegraphics[width=0.49\columnwidth]{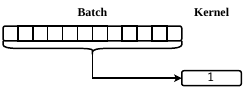}%
	}}%
	\subfloat[Multiple Instances\label{subfig:multi}]{
		\includegraphics[width=0.49\columnwidth]{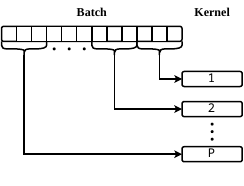}
	}%
	\caption{Batch distribution for single and multiple kernel instances.}
	\label{fig:mulitKernel}
\end{figure}

The proposed design is resource-efficient enough to be able to implement multiple kernel instances on the same device. This opens up further optimization potential by executing identical search kernels in parallel on the FPGA, given that multiple memory controllers are available. This concept is illustrated in Fig.~\ref{subfig:multi}, where \texttt{P} denotes the parallelization parameter.

The batch of search keys can be evenly distributed based on the parallelization factor \texttt{P}, allowing either a larger number of keys to be processed simultaneously or a reduction in execution time for the same batch size.

\section{Evaluation} \label{sec:evaluation}

The presented FPGA design for the B+ tree search algorithm was developed using the AMD Vitis high-level synthesis (HLS) flow. The system architecture consists of a CPU and an FPGA connected via PCIe. In the following sections, the system architecture is described in detail, and the performance of the FPGA-based design is analyzed and compared to a CPU-based B+ tree search algorithm.

\subsection{System Architecture}

\begin{figure}[h!]
	\centering
	\includegraphics[width=0.8\columnwidth]{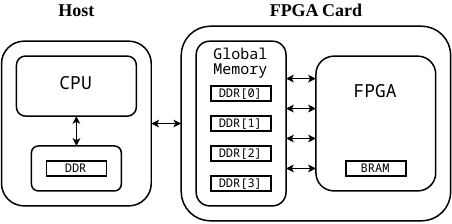}
	\caption{System Architecture}
	\label{fig:architecture}
\end{figure}

An overview of the hybrid system architecture is provided in Fig.~\ref{fig:architecture}, which consists of two main components: the host system and the FPGA board. The host system includes a CPU with its own DDR memory, while the FPGA board comprises programmable logic (PL) and four DDR memory modules, each with a capacity of \SI{16}{\giga\byte}. The DDR memory on the FPGA board is referred to as global memory, which can be accessed by the CPU via PCIe and by the PL via AXI Stream interfaces. Specifically, the host system is equipped with an AMD EPYC 7542 processor running at a base frequency of \SI{2.9}{\giga\hertz} and a turbo frequency of up to \SI{3.4}{\giga\hertz}. The FPGA platform used is an AMD Alveo U250 Data Center Accelerator Card operating at \SI{300}{\mega\hertz}.

As the focus of this work is on accelerating a static B+ tree, the index structure is maintained on the host system. For this purpose, we utilize the B+ tree implementation provided by the TLX Library~\cite{Bingmann2018}. The B+ trees are generated using the TLX library and we employ host-side code to transform the tree structure into flat arrays suitable for transfer to the FPGA's global memory. Based on formula~\eqref{eq:node_size} and the available global memory, trees with up to 1.6 billion entries can be stored, corresponding to approximately 25 million nodes for $m=64$ and up to 100 million nodes for $m=16$. Additionally, we generate random tree entries and search keys to build the B+ tree and corresponding query batches. To verify the correctness of the FPGA-based search, the same query batch is evaluated using the TLX search function, and the results are compared.

\subsection{Timing Methodology} \label{subsec:timing}

The timing measurement of a single search request can be divided into two main phases. The first phase encompasses the initialization steps performed on the host side, including the construction and transformation of the B+ tree, the OpenCL setup, the FPGA programming, and the transfer of the tree structure to the global memory. These steps are part of a one-time initialization process and are not repeated for every new search request. For each subsequent query, only the new batch of search keys needs to be transferred to the FPGA.

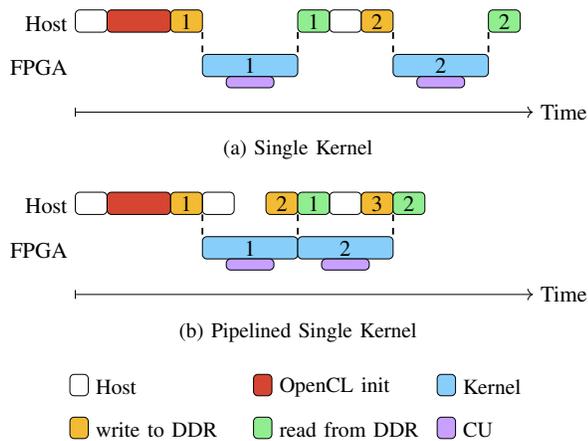
\begin{figure}[]
	\centering
	\subfloat[Single Kernel\label{subfig:singleTiming}]{
		\resizebox{0.9\columnwidth}{!}{%
			\begin{tikzpicture}[x=0.1cm, y=0.7cm, rounded corners=2pt]
				
				\definecolor{openCL}{RGB}{214, 69, 49}    
				\definecolor{wDDR}{RGB}{243, 187, 51}     
				\definecolor{rDDR}{RGB}{144, 238, 144}    
				\definecolor{kernel}{RGB}{135, 206, 250}  
				\definecolor{CU}{RGB}{200, 160, 255}      
				
				\draw[->] (0, -3) -- (72, -3) node[right] {Time};
				\draw (0, -3.1) -- (0, -2.9);
				
				\foreach \bulkName/\i in {Host/1, FPGA/2} {
					\coordinate (\bulkName\i) at (0, -\i-0.2);
					
					\node[anchor=east] at (0, -\i) {\bulkName};
				}
				
				\draw[] (Host1) rectangle ++(5, 0.5);
				\coordinate (Host1) at ($(Host1) + (5, 0)$);
				\draw[fill=openCL] (Host1) rectangle ++(10, 0.5);
				\coordinate (Host1) at ($(Host1) + (10, 0)$);
				\draw[fill=wDDR] (Host1) rectangle ++(5, 0.5);
				\node[anchor=center] at ($(Host1) + (2.5, 0.25)$) {1};
				\coordinate (Kernel1) at ($(Host1) + (5, 0)$);
				\coordinate (Host2) at ($(Host1) + (5, 0)$);
				\coordinate (Host2) at ($(Host2) + (10, 0)$);
				
				\coordinate (FPGA2) at ($(Kernel1) + (0, -1)$);
				\draw[fill=kernel] (FPGA2) rectangle ++(15, 0.5);
				\node[anchor=center] at ($(FPGA2) + (7.5, 0.25)$) {1};
				\draw[fill=CU] ($(FPGA2) + (3.75, -0.25)$) rectangle ++(7.5, 0.25);
				
				\coordinate (Host3) at ($(Host2) + (5, 0)$);
				\draw[fill=rDDR] (Host3) rectangle ++(5, 0.5);
				\node[anchor=center] at ($(Host3) + (2.5, 0.25)$) {1};
				\coordinate (Host3) at ($(Host3) + (5, 0)$);
				\draw[] (Host3) rectangle ++(5, 0.5);
				\coordinate (Host3) at ($(Host3) + (5, 0)$);
				\draw[fill=wDDR] (Host3) rectangle ++(5, 0.5);
				\node[anchor=center] at ($(Host3) + (2.5, 0.25)$) {2};
				
				\coordinate (FPGA3) at ($(FPGA2) + (30, 0)$);
				\draw[fill=kernel] (FPGA3) rectangle ++(15, 0.5);
				\node[anchor=center] at ($(FPGA3) + (7.5, 0.25)$) {2};
				\draw[fill=CU] ($(FPGA3) + (3.75, -0.25)$) rectangle ++(7.5, 0.25);
				
				\coordinate (Host4) at ($(Host3) + (20, 0)$);
				\draw[fill=rDDR] (Host4) rectangle ++(5, 0.5);
				\node[anchor=center] at ($(Host4) + (2.5, 0.25)$) {2};

				\draw[dashed, thick] (Kernel1) -- ($(Kernel1) + (0, -0.5)$);
				\draw[dashed, thick] ($(Kernel1) + (15, 0)$) -- ($(Kernel1) + (15, -0.5)$);
				
				\draw[dashed, thick] ($(FPGA3) + (0, 1)$) -- ($(FPGA3) + (0, 0.5)$);
				\draw[dashed, thick] ($(FPGA3) + (15, 1)$) -- ($(FPGA3) + (15, 0.5)$);
				
			\end{tikzpicture}
	}}%
	\\
	\subfloat[Pipelined Single Kernel\label{subfig:idealSingleTiming}]{
		\resizebox{0.9\columnwidth}{!}{%
			\begin{tikzpicture}[x=0.1cm, y=0.7cm, rounded corners=2pt]
				
				\definecolor{openCL}{RGB}{214, 69, 49}    
				\definecolor{wDDR}{RGB}{243, 187, 51}     
				\definecolor{rDDR}{RGB}{144, 238, 144}    
				\definecolor{kernel}{RGB}{135, 206, 250}  
				\definecolor{CU}{RGB}{200, 160, 255}      
				
				\draw[->] (0, -3) -- (72, -3) node[right] {Time};
				\draw (0, -3.1) -- (0, -2.9);
				
				\foreach \bulkName/\i in {Host/1, FPGA/2} {
					\coordinate (\bulkName\i) at (0, -\i-0.2);
					
					\node[anchor=east] at (0, -\i) {\bulkName};
				}
				
				\draw[] (Host1) rectangle ++(5, 0.5);
				\coordinate (Host1) at ($(Host1) + (5, 0)$);
				\draw[fill=openCL] (Host1) rectangle ++(10, 0.5);
				\coordinate (Host1) at ($(Host1) + (10, 0)$);
				\draw[fill=wDDR] (Host1) rectangle ++(5, 0.5);
				\node[anchor=center] at ($(Host1) + (2.5, 0.25)$) {1};
				\coordinate (Kernel1) at ($(Host1) + (5, 0)$);
				\coordinate (Host2) at ($(Host1) + (5, 0)$);
				\draw[] (Host2) rectangle ++(5, 0.5);
				\coordinate (Host2) at ($(Host2) + (10, 0)$);
				\draw[fill=wDDR] (Host2) rectangle ++(5, 0.5);
				\node[anchor=center] at ($(Host2) + (2.5, 0.25)$) {2};
				
				\coordinate (FPGA2) at ($(Kernel1) + (0, -1)$);
				\draw[fill=kernel] (FPGA2) rectangle ++(15, 0.5);
				\node[anchor=center] at ($(FPGA2) + (7.5, 0.25)$) {1};
				\draw[fill=CU] ($(FPGA2) + (3.75, -0.25)$) rectangle ++(7.5, 0.25);
				
				\coordinate (Host3) at ($(Host2) + (5, 0)$);
				\draw[fill=rDDR] (Host3) rectangle ++(5, 0.5);
				\node[anchor=center] at ($(Host3) + (2.5, 0.25)$) {1};
				\coordinate (Host3) at ($(Host3) + (5, 0)$);
				\draw[] (Host3) rectangle ++(5, 0.5);
				\coordinate (Host3) at ($(Host3) + (5, 0)$);
				\draw[fill=wDDR] (Host3) rectangle ++(5, 0.5);
				\node[anchor=center] at ($(Host3) + (2.5, 0.25)$) {3};
				
				\coordinate (FPGA3) at ($(FPGA2) + (15, 0)$);
				\draw[fill=kernel] (FPGA3) rectangle ++(15, 0.5);
				\node[anchor=center] at ($(FPGA3) + (7.5, 0.25)$) {2};
				\draw[fill=CU] ($(FPGA3) + (3.75, -0.25)$) rectangle ++(7.5, 0.25);
				
				\coordinate (Host4) at ($(Host3) + (5, 0)$);
				\draw[fill=rDDR] (Host4) rectangle ++(5, 0.5);
				\node[anchor=center] at ($(Host4) + (2.5, 0.25)$) {2};
				

				\draw[dashed, thick] (Kernel1) -- ($(Kernel1) + (0, -0.5)$);
				\draw[dashed, thick] ($(Kernel1) + (15, 0)$) -- ($(Kernel1) + (15, -0.5)$);
				\draw[dashed, thick] ($(FPGA3) + (15, 1)$) -- ($(FPGA3) + (15, 0.5)$);
				
			\end{tikzpicture}
	}}%
	\\
	\subfloat{
		\resizebox{0.7\columnwidth}{!}{%
			\begin{tikzpicture}[x=0.1cm, y=0.7cm, rounded corners=2pt]
				
				\definecolor{openCL}{RGB}{214, 69, 49}    
				\definecolor{wDDR}{RGB}{243, 187, 51}     
				\definecolor{rDDR}{RGB}{144, 238, 144}    
				\definecolor{kernel}{RGB}{135, 206, 250}  
				\definecolor{CU}{RGB}{200, 160, 255}      
				
				\begin{scope}[xshift=0cm, yshift=0cm]
					\coordinate (legend-start) at (0, 0);
					
					\draw[rounded corners=2pt] (legend-start) rectangle ++(3, 0.5);
					\node[anchor=west] at ($(legend-start) + (3, 0.2)$) {Host};
					
					\coordinate (legend-2) at ($(legend-start) + (30, 0)$);
					\draw[fill=openCL, rounded corners=2pt] (legend-2) rectangle ++(3, 0.5);
					\node[anchor=west] at ($(legend-2) + (3, 0.2)$) {OpenCL init};
					
					\coordinate (legend-5) at ($(legend-start) + (60, 0)$);
					\draw[fill=kernel, rounded corners=2pt] (legend-5) rectangle ++(3, 0.5);
					\node[anchor=west] at ($(legend-5) + (3, 0.2)$) {Kernel};
					
					\coordinate (legend-3) at ($(legend-start) + (0, -1)$);
					\draw[fill=wDDR, rounded corners=2pt] (legend-3) rectangle ++(3, 0.5);
					\node[anchor=west] at ($(legend-3) + (3, 0.2)$) {write to DDR};
					
					\coordinate (legend-4) at ($(legend-start) + (30, -1)$);
					\draw[fill=rDDR, rounded corners=2pt] (legend-4) rectangle ++(3, 0.5);
					\node[anchor=west] at ($(legend-4) + (3, 0.2)$) {read from DDR};
					
					\coordinate (legend-6) at ($(legend-start) + (60, -1)$);
					\draw[fill=CU, rounded corners=2pt] (legend-6) rectangle ++(3, 0.5);
					\node[anchor=west] at ($(legend-6) + (3, 0.2)$) {CU};
				\end{scope}	
				
			\end{tikzpicture}
	}}%
	\caption{Interaction between Host and FPGA.}
	\label{fig:timingDiagram}
	\vspace{-10pt}
\end{figure}

The second phase begins with the kernel execution on the FPGA. This includes the kernel initialization as well as the Compute Unit (CU) execution, where the actual search operation takes place. The dependencies and the temporal flow of these steps are illustrated in Fig.~\ref{subfig:singleTiming}. Before initiating the next request, the host system must read the previous results, generate a new batch of search queries, and write this batch to the global memory.

The pipelined timing scenario is illustrated in Fig.~\ref{subfig:idealSingleTiming}. The host can overlap data transfers and result collection with ongoing kernel execution. This is feasible because the search keys are preloaded into the FPGA’s BRAM, eliminating the need for global memory access during computation. For a sufficiently large batch size data transfers and host-side operations can be fully masked by the FPGA’s kernel execution time and no additional timing overhead occurs compared to a pure CPU-based execution.

For our measurements, we use a system with a static B+ tree index structure that is transferred to the FPGA once and reused thereafter. \added{Therefore, }the overhead of this one-time initialization phase becomes negligible in a productive system. The experimental results are obtained using the flow shown in Fig.~\ref{subfig:singleTiming}. For all results, we report the the kernel and compute unit execution times according to Fig.~\ref{subfig:idealSingleTiming}, as these are the decisive execution times.


\subsection{Experimental Setup}

To conduct a comprehensive analysis of the proposed design, we analyzed multiple test configurations. For the B+ tree construction, we vary the tree size (the number of entries in the tree), the tree order $m$, and the batch size (the number of search keys processed in one iteration). The tree size is independent of the hardware design and is limited only by the available DDR memory on the FPGA board. In our evaluation, we selected tree sizes ranging from 1 to 10 million entries. The tree order must be set prior to synthesis, as it defines the node size and determines structural aspects of the comparison logic. Consequently, we implemented separate hardware designs for three different tree orders: $m = 16$, $m = 32$, and $m = 64$. The batch size is configurable at runtime, with a maximum of 1000 queries, and can be varied within this range during testing.

In general, we conduct timing measurements for both the FPGA and CPU implementations of the search algorithm. For the FPGA, we integrate trace logic capable of recording timestamps for kernel execution, compute unit activity, and OpenCL API calls. On the host side, timing is measured using the \texttt{chrono} module from the standard C++ library. 


To obtain more meaningful and robust results, each configuration is repeated multiple times: 10 iterations for the FPGA measurements and 100 for the CPU measurements. Due to occasional significant outliers in the CPU timings, we report the results using the interquartile mean $x_{\text{IQM}}$. This metric discards the lower and upper quartiles (i.e., the lowest and highest 25\% of values) and computes the arithmetic mean over the central 50\% of the data~\cite{Dodge2008}:

\begin{equation}
	x_{\text{IQM}} = \frac{1}{n} \cdot \sum_{i=\frac{n}{4}+1}^{\frac{3n}{4}} x_i
\end{equation}

Additionally, we use the interquartile range (IQR), the difference between the upper and lower quartiles, to capture variability~\cite{Dodge2008}. In the following plots, the IQR is visualized as error bars, indicating the range within which the central half of the data lies.

\subsection{Resource Utilization}

The FPGA resource utilization is summarized in Tab.~\ref{tab:resources} for both single and four-kernel instance designs. To contextualize the results, the table provides both the absolute number of FPGA resources used (\texttt{Total}) and the available resources (\texttt{Available}) for individual designs. Their difference reflects the resource utilization consumed by the FPGA shell and the system.

\textbf{Single Instance Utilization.} For the single instance designs, it is evident that the maximum achievable frequency is maintained for the configurations with $m=16$ and $m=32$. In contrast, the design with $m=64$ achieves only approximately two-third of the maximum frequency, indicating that higher-order trees introduce timing challenges due to the increased parallel logic in the search and comparison blocks. All three designs demonstrate very efficient resource usage, utilizing only 0.74-1.71\% of the available Look-Up Tables (LUTs) and 0.33-0.76\% of the Flip-Flops (FFs). The moderate increase in resource usage from $m=16$ to $m=64$ is a direct result of the design structure. As the number of tree orders increases to $m=32$ and $m=64$, the utilization of LUTs increases by factors of 1.5$\times$ and 1.8$\times$, respectively.
The Block-RAM and Digital Signal Processing (DSP) unit utilization remains constant across all configurations. This consistency can be attributed to the fact that the allocated memory for search keys and result buffers does not vary with the tree order. Only the size of the currently loaded node changes, implying that node data is stored using distributed RAM, rather than BRAM.

\textbf{Four Instances Utilization.} The resource utilization of the four-instance design scales similarly to the single-instance design, as the kernel is simply instantiated four times, leading to a roughly fourfold increase in overall resource usage. The design is generally timing-bound. The configurations with $m=16$ and $m=32$ are close to the target frequency and the synthesis fails with timing violations for the configuration with $m=64$ since no further modifications are made to the search kernel.
\begin{table*}[h!]
	\centering
	\caption{FPGA Resource Utilization}
	\begin{tabular}{rr|c|rr|rr|rr|rr|rr}
		& &
		\multirow{2}{1.4cm}{Clock Freq. in \si{\mega\hertz}}	& 
		\multirow[t]{2}{1cm}{\centering LUT} 					& \multirow{2}{0.6cm}{Util. in \%} & 
		\multirow{2}{1cm}{LUT as Mem}    						& \multirow{2}{0.6cm}{Util. in \%} & 
		\multirow[t]{2}{1cm}{\centering FF} 					& \multirow{2}{0.6cm}{Util. in \%} & 
		\multirow[t]{2}{1cm}{\centering BRAM} 					& \multirow{2}{0.6cm}{Util. in \%} & 
		\multirow[t]{2}{1cm}{\centering DSP} 					& \multirow{2}{0.6cm}{Util. in \%} \\
		& & & & & & & & & & & & \\
		\hline\hline
		& Total 		&    & \SI{1726208}{} &  			& \SI{790192}{} &  				& \SI{3456000}{} &  			& \SI{2688}{} &  			& \SI{12288}{} & \\ 
		\hline
		& Available	&  			   & \SI{1624067}{} & \SI{94.08}{} & \SI{769486}{} & \SI{97.38}{} 	& \SI{3291026}{} & \SI{95.23}{} & \SI{2282}{} & \SI{84.9}{} & \SI{12284}{} & \SI{99.97}{} \\ 
		\multirow[c]{3}{*}{\rotatebox{90}{1 Inst.}} & $m=16$ & \SI{300.0}{} &   \SI{11991}{} 	 & \SI{0.74}{} 	& \SI{1090}{} 	& \SI{0.14}{} 	& \SI{10861}{}   & \SI{0.33}{} 	& \SI{22}{}   & \SI{0.96}{} & \SI{0}{} 	   & \SI{0}{} \\ 
		& $m=32$ & \SI{300.0}{} &   \SI{15374}{} 	 & \SI{0.95}{} 	& \SI{1098}{} 	& \SI{0.14}{} 	& \SI{15695}{}   & \SI{0.48}{} 	& \SI{22}{}   & \SI{0.96}{} & \SI{0}{} 	   & \SI{0}{} \\ 
		& $m=64$ & \SI{199.4}{} &   \SI{27750}{} & \SI{1.71}{} 	& \SI{1344}{} 	& \SI{0.17}{} 	& \SI{24954}{}   & \SI{0.76}{} 	& \SI{22}{}   & \SI{0.96}{} & \SI{0}{} 	   & \SI{0}{} \\ 
		\hline
		\multirow[c]{10}{*}{\rotatebox{90}{4 Instances}} & Available 	&    & \SI{1522722}{} & \SI{88.21}{} & \SI{754684}{} & \SI{95.51}{} & \SI{3139326}{} & \SI{90.84}{} & \SI{2068}{} & \SI{76.93}{} & \SI{12275}{} & \SI{99.89}{} \\ 
		& $m=16$ & \SI{293.2}{}   & \SI{47576}{} & \SI{3.12}{} & \SI{4360}{} & \SI{0.58}{} & \SI{43431}{} & \SI{1.38}{} & \SI{88}{} & \SI{4.26}{} & \SI{0}{} & \SI{0}{} \\ 
		& 1. Inst. &    & \SI{11885}{} & \SI{0.78}{} & \SI{1090}{} & \SI{0.14}{} & \SI{10864}{} & \SI{0.35}{} & \SI{22}{} & \SI{1.06}{} & \SI{0}{} & \SI{0}{} \\ 
		& 2. Inst. &    & \SI{11866}{} & \SI{0.78}{} & \SI{1090}{} & \SI{0.14}{} & \SI{10825}{} & \SI{0.34}{} & \SI{22}{} & \SI{1.06}{} & \SI{0}{} & \SI{0}{} \\ 
		& 3. Inst. &    & \SI{11870}{} & \SI{0.78}{} & \SI{1090}{} & \SI{0.14}{} & \SI{10869}{} & \SI{0.35}{} & \SI{22}{} & \SI{1.06}{} & \SI{0}{} & \SI{0}{} \\ 
		& 4. Inst. &    & \SI{11955}{} & \SI{0.79}{} & \SI{1090}{} & \SI{0.14}{} & \SI{10873}{} & \SI{0.35}{} & \SI{22}{} & \SI{1.06}{} & \SI{0}{} & \SI{0}{} \\ 
		& $m=32$ & \SI{300.0}{}   & \SI{61400}{} & \SI{4.03}{} & \SI{4392}{} & \SI{0.58}{} & \SI{62668}{} & \SI{2.00}{} & \SI{88}{} & \SI{4.26}{} & \SI{0}{} & \SI{0}{} \\ 
		& 1. Inst. &    & \SI{15331}{} & \SI{1.01}{} & \SI{1098}{} & \SI{0.15}{} & \SI{15711}{} & \SI{0.50}{} & \SI{22}{} & \SI{1.06}{} & \SI{0}{} & \SI{0}{} \\ 
		& 2. Inst. &    & \SI{15336}{} & \SI{1.01}{} & \SI{1098}{} & \SI{0.15}{} & \SI{15602}{} & \SI{0.50}{} & \SI{22}{} & \SI{1.06}{} & \SI{0}{} & \SI{0}{} \\ 
		& 3. Inst. &    & \SI{15370}{} & \SI{1.01}{} & \SI{1098}{} & \SI{0.15}{} & \SI{15671}{} & \SI{0.50}{} & \SI{22}{} & \SI{1.06}{} & \SI{0}{} & \SI{0}{} \\ 
		& 4. Inst. &    & \SI{15363}{} & \SI{1.01}{} & \SI{1098}{} & \SI{0.15}{} & \SI{15684}{} & \SI{0.50}{} & \SI{22}{} & \SI{1.06}{} & \SI{0}{} & \SI{0}{} \\
		\hline
	\end{tabular}
	\label{tab:resources}
	\vspace{-10pt}
\end{table*}

\subsection{FPGA Performance}

In the following, we present the measured timing results for the FPGA designs. These measurements were conducted as described in Sec.\ref{subsec:timing} and Fig.\ref{subfig:singleTiming}.

\begin{figure}
	\centering
	\resizebox{0.9\columnwidth}{!}{\batchCase{\input{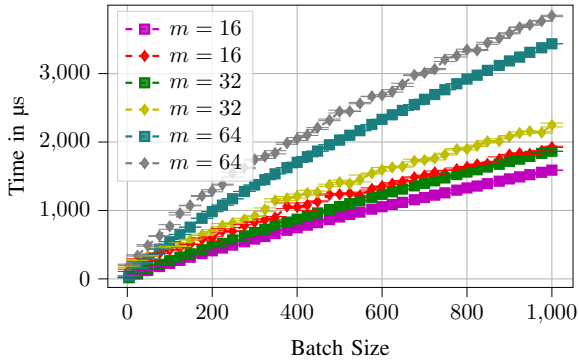}}}
	\caption{\textbf{Single Instance Designs} - CU (\tikzmarker{square*}) and Kernel (\tikzmarker{diamond*}) execution times with a fix tree size of 1 million entries. Each configuration is 10$\times$ repeated.}
	\label{fig:1_fpga_batchCase_1Mio_all}
	\vspace{-10pt}
\end{figure}

\textbf{Single Instance Designs.} Fig.~\ref{fig:1_fpga_batchCase_1Mio_all} presents the Kernel and Compute Unit execution times for the three single-instance FPGA designs. The tree size is fixed at one million entries, while the batch size \replaced{varies from 1}{is varied up} to 1000 search keys. There is a constant overhead of approximately \SI{150}{\micro\second} between the kernel and CU times. The CU time reflects the duration for transferring nodes between global memory and the FPGA, as well as the duration for executing all search operations for the given batch. As expected, the time per search key decreases as batch sizes increase.

As the batch size increases, more comparisons are required, and additional tree nodes must be loaded from global memory. This leads to a corresponding increase in execution times across all configurations. The notably slower performance observed for the tree order $m=64$ is due to the fact that this design operates at only two-thirds of the maximum achievable clock frequency. Overall, the smallest tree order, $m=16$, consistently delivers the best performance across all batch sizes.

\begin{figure}[]
	\centering
	\resizebox{0.9\columnwidth}{!}{\batchCaseSmallCU{\input{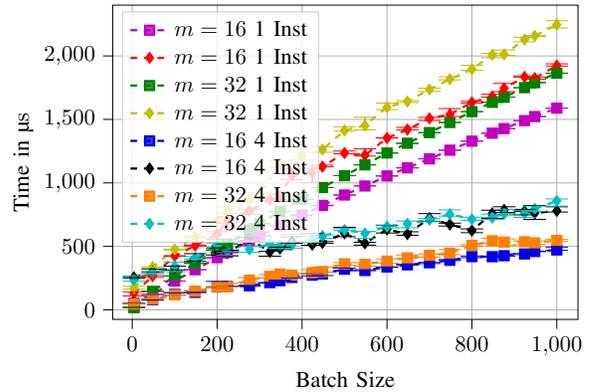}}}
	\caption{\textbf{Four Instance Designs} - CU (\tikzmarker{square*}) and Kernel (\tikzmarker{diamond*}) execution times with a fix tree size of 1 million entries. Each configuration is 10$\times$ repeated.}
	\label{fig:4_fpga_batchCase_1Mio_15_31}
	\vspace{-10pt}
\end{figure}

\textbf{Four Instance Designs.} Beside the three baseline designs, we evaluate two optimized configurations in which the same kernel is instantiated four times, with each instance assigned to a dedicated DDR memory bank. The results for tree orders $m=16$ and $m=32$ are shown in Fig.~\ref{fig:4_fpga_batchCase_1Mio_15_31}. These results demonstrate that kernel-level parallelism and batch distribution across four independent kernels, each obtaining 250 search keys, significantly reduce the kernel execution time. 



With a batch size of 1000, the CU execution times for the single- and four-instance designs show a speedup of 3.4$\times$, which approaches the ideal 4$\times$ acceleration. In terms of kernel execution time, speedups of 2.5$\times$ and 2.6$\times$ are achieved for tree orders $m=16$ and $m=32$, respectively. These lower speedups compared to the CU times are primarily due to kernel initialization and setup overhead.

\subsection{CPU and FPGA Performance Comparison}

In the following part, the presented FPGA results are compared against a pure CPU execution using the search function provided by the TLX Library~\cite{Bingmann2018}. The search algorithm in this work is specifically designed for FPGA execution and performs worse on a pure CPU compared to the TLX search algorithm.

\begin{figure}
	\centering
	\resizebox{0.9\columnwidth}{!}{\batchCaseCPU{\input{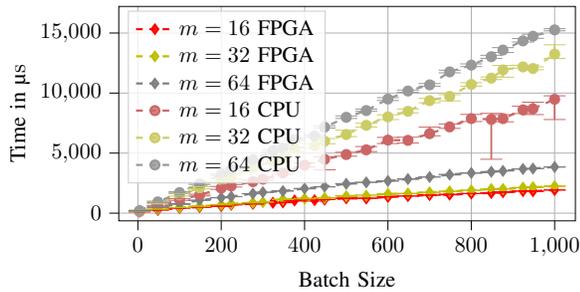}}}
	\caption{\textbf{Single-Threaded CPU} - FPGA kernel (\tikzmarker{diamond*}) and CPU search (\tikzmarker{*}) execution times with a fix tree size of 1 million entries. FPGA configuration is 10$\times$ and CPU is 100$\times$ repeated.}
	\label{fig:fpga_CPU_batchCase_1Mio_all}
	\vspace{-5pt}
\end{figure}

\textbf{Single-Threaded CPU.} Fig.~\ref{fig:fpga_CPU_batchCase_1Mio_all} shows the kernel execution times for all three tree orders measured on the FPGA, alongside the corresponding search execution times for the same batches on a single-threaded CPU. The FPGA consistently outperforms the CPU across all configurations. The single instance FPGA design achieves with $m=16$ a speedup of 4.9$\times$ compared to the single threaded CPU search. The CPU measurements exhibit high variability, which may be due to the task not having exclusive access to the CPU core.

It can be concluded that the FPGA and CPU implementations achieve optimal performance with a tree order of $m=16$. Therefore, only the $m=16$ tree order will be considered in the following.

\begin{figure}
	\centering
	\resizebox{0.9\columnwidth}{!}{\batchCaseCPUThread{\input{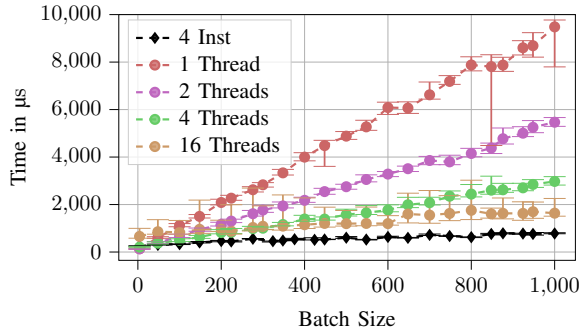}}}
	\caption{\textbf{Multi-Threaded CPU} - FPGA kernel (\tikzmarker{diamond*}) and CPU search (\tikzmarker{*}) execution times with a fix tree size of 1 million entries and a tree order of $m=16$. FPGA configuration is 10$\times$ and CPU is 100$\times$ repeated.}
	\label{fig:fpga_CPU_multi_batchCase_1Mio_all}
	\vspace{-10pt}
\end{figure}

\textbf{Multi-Threaded CPU.} Database queries on CPUs are typically distributed across multiple threads. We analyze multi-threaded CPU performance in Fig.~\ref{fig:fpga_CPU_multi_batchCase_1Mio_all}. Following the same strategy as with the four-instance FPGA design, the batch is evenly divided among the available CPU threads. We vary the number of threads up to 16, observing a saturation point beyond which additional threads yield minimal performance gains. At very small batch sizes, the multi-threaded CPU execution performs worse than the single-threaded case due to synchronization overhead. However, this effect becomes negligible as the batch size increases. The maximum speedup achieved by the four-instance FPGA design over the 16-thread CPU configuration is 2.1$\times$ at a batch size of 1000.

\begin{figure}[h!]
	\centering
	\resizebox{0.9\columnwidth}{!}{\treeCase{\input{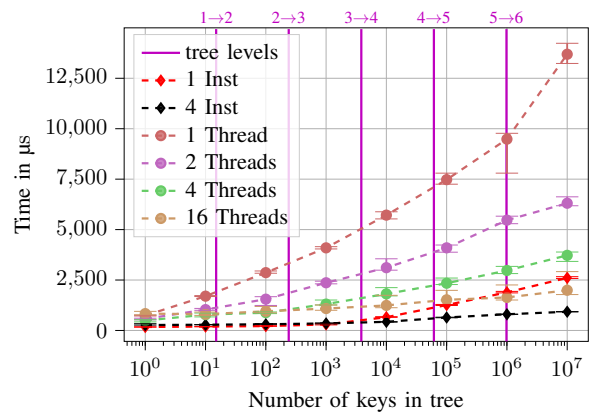}}}
	\caption{\textbf{Varying Tree Sizes} - FPGA kernel (\tikzmarker{diamond*}) and CPU search (\tikzmarker{*}) execution times with a fix batch size of 1000 search keys and a tree order of $m=16$. FPGA configuration is 10$\times$ and CPU is 100$\times$ repeated.}
	\label{fig:fpga_CPU_treeCase_1000_16}
	\vspace{-15pt}
\end{figure}

\textbf{Varying Tree Sizes.} Fig.~\ref{fig:fpga_CPU_treeCase_1000_16} shows the execution time for varying tree sizes, ranging from 1 entry to 10 million entries, using a fixed batch size of 1000. The 16-threaded CPU configuration outperforms the single-instance FPGA implementation for trees larger than 100,000 entries. However, the four-kernel FPGA design still outperforms the 16-threaded CPU version at a tree size of 1 million entries. The results for tree sizes larger than 1 million follow the same trend, with slight performance improvements. At 10 million entries, a speedup of 2.15$\times$ is achieved.

The slightly better performance of the single-instance FPGA design at small tree sizes compared to the four-instance variant is due to the fact that the kernel instances are not launched at exactly the same time; the start times of the instances are  shifted, which slightly increases the reported execution time. In summary, all results of the evaluation are presented in Table~\ref{tab:speedup}.
\begin{table}[h!]
	\centering
	\caption{Speedup Overview}
	\begin{tabular}{c|c|c|c|c}
		& \textbf{Tree Order} &\textbf{Comparison$^{\mathrm{1}}$} & \textbf{CU} & \textbf{Kernel} \\
		\hline
		\multirow[c]{2}{*}{FPGA} 
		& $m=16$ & \multicolumn{1}{l|}{1 Inst. with 4 Inst.} & 3.4$\times$ & 2.5$\times$ \\
		& $m=32$ & \multicolumn{1}{l|}{1 Inst. with 4 Inst.} & 3.4$\times$ & 2.6$\times$ \\
		\hline
		\multirow[c]{2}{1cm}{\centering FPGA\\\& CPU} 
		& $m=16$ & \multicolumn{1}{l|}{1 Inst. with 1 Thread}	& - & 4.9$\times$ \\
		& $m=16$ & \multicolumn{1}{l|}{4 Inst. with 16 Threads} & - & 2.1$\times$ \\
		\multicolumn{5}{l}{$^{\mathrm{1}}$based on batch size of 1000 and tree size of 1 million}\\
	\end{tabular}
	\label{tab:speedup}
	\vspace{-5pt}
\end{table}



\section{Conclusion} \label{sec:conclusion}


The evaluation demonstrates significant performance improvements for B+ tree search on the AMD Alveo U250 Data Center Accelerator Card. The proposed approach achieves a speedup of up to 4.9$\times$ for the single-instance design and 2.1$\times$ for the four-instance design compared to single- and multi-threaded CPU implementations, respectively. These gains are enabled by a dedicated hardware-optimized search algorithm that effectively exploits the parallelism of the FPGA while improving memory access efficiency.

The results further show that the proposed design is highly scalable and particularly effective for static B+ trees. This characteristic makes it well suited for use as an index structure for caching hot subsets in data warehouse environments, where data remains largely unchanged and query performance is critical.

Overall, the design utilizes only a small fraction of the available hardware resources while maintaining scalability with increasing tree sizes. As the tree grows, the execution time per search operation decreases, highlighting the efficiency of the approach for large-scale workloads.

\bibliographystyle{bib/IEEEtran}
\bibliography{bib/IEEEabrv,bib/fpt_paper_lib}
\clearpage

\end{document}